\documentclass{article}
\usepackage[T1]{fontenc}
\usepackage[utf8]{inputenc}
\usepackage{spconf}
\usepackage{bm,amsmath,cite,url,pifont,amssymb, booktabs}
\usepackage{multirow, multicol}
\usepackage{graphicx}
\usepackage{color}
\usepackage{acronym}
\usepackage[%
 setpagesize=false,%
 bookmarks=true,%
 bookmarksdepth=tocdepth,%
 bookmarksnumbered=true,%
 colorlinks=false,%
 pdftitle={},%
 pdfsubject={},%
 pdfauthor={},%
 pdfkeywords={}%
]{hyperref}

\title{Annotation-free MIDI-to-Audio Synthesis via Concatenative Synthesis and Generative Refinement}

\name{Osamu Take and Taketo Akama} %\thanks{Thanks to XYZ agency for funding.}
\address{Sony Computer Science Laboratories, Tokyo, Japan}

\newcommand{\bR}{\mathbb{R}}
\newacro{generative refinement module}[Ref]{generative refinement module}
\newacro{concatenative sampler}[CoSa]{concatenative sampler}

\begin{document}

\maketitle

\begin{abstract}
Recent MIDI-to-audio synthesis methods using deep neural networks have successfully generated high-quality, expressive instrumental tracks. 
However, these methods require MIDI annotations for supervised training, limiting the diversity of instrument timbres and expression styles in the output.
We propose \textit{CoSaRef}, a MIDI-to-audio synthesis method that does not require MIDI-audio paired datasets. CoSaRef first generates a synthetic audio track using concatenative synthesis based on MIDI input, then refines it with a diffusion-based deep generative model trained on datasets without MIDI annotations. 
This approach improves the diversity of timbres and expression styles. Additionally, it allows detailed control over timbres and expression through audio sample selection and extra MIDI design, similar to traditional functions in digital audio workstations.
Experiments showed that CoSaRef could generate realistic tracks while preserving fine-grained timbre control via one-shot samples. 
Moreover, despite not being supervised on MIDI annotation, CoSaRef outperformed the state-of-the-art timbre-controllable method based on MIDI supervision in both objective and subjective evaluation.
\end{abstract}

\section{Introduction}
\label{sec:intro}
Nowadays, for emulating real-world instrument performances, music creators frequently use musical instruments such as sample-concatenative~\cite{Schwarz01032006} and physical-modeling~\cite{physical-modeling} synthesizers to generate tracks. 
These widely used instruments perform MIDI-to-audio synthesis by taking MIDI representations~\cite{midi}, including note onset and offset, duration, pitch, and velocity, as input and producing audio by rendering the input MIDI scores.
While they can produce high-quality audio at the note level, they struggle to capture the expressive nuances and realism in human performances at the phrase and track levels.
For instance, musicians naturally incorporate techniques such as slurs and slides in violins or breathiness in woodwinds, shaping the expressiveness of a performance.
Conventional synthesizers often fail or require great effort from users to reproduce these intricate details, limiting the realism of the generated audio.

Recent studies have incorporated deep generative models for MIDI-to-audio synthesis. 
Thanks to their data-driven approach, these models have enabled musicians to generate more expressive, human-like instrumental tracks compared to conventional synthesis methods. 
Instrumental audio datasets with MIDI annotations, such as MusicNet~\cite{musicnet}, have facilitated the development of deep generative MIDI-to-audio synthesis in a supervised manner.

This supervised learning-based approach inevitably faces challenges related to data collection. 
Training data for supervised MIDI-to-audio synthesis requires precise time alignment of each note for each realistic performance.
However, there are only a few open-source instrumental audio datasets that meet this time-aligned MIDI requirement, and they are limited in timbre diversity, including instrument types and performance styles.
Creating new datasets to address this issue should be a challenging task. 
The limited performance of automatic music transcription and MIDI alignment to instrumental audio makes it hard to automate some dataset construction processes.
Humans can annotate the MIDI transcription and note onset/offset accurately, while this process is costly. 

\begin{figure*}[t]
  \centering
  \includegraphics[width=0.8\linewidth]{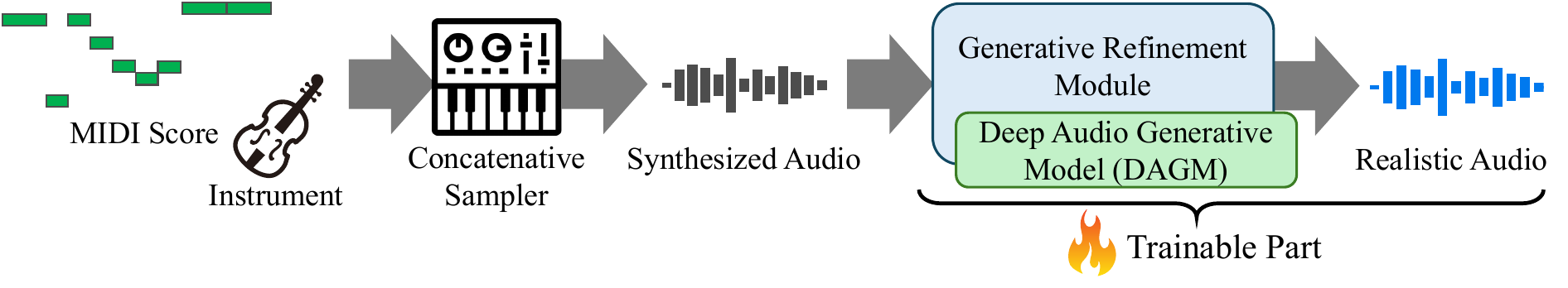}
%  \vspace{2.0cm}
%
\vspace{-6mm}
\caption{Diagram of the CoSaRef's framework for expressive MIDI-to-audio synthesis. 
%The trainable part of this framework is DAGM, which requires only the unpaired audio collection for training.
}
\label{fig:darefine-demo}
\vspace{-3mm}
\end{figure*}
This study introduces \textbf{Co}ncatenative \textbf{Sa}mpler and \textbf{Ref}inement (\textit{CoSaRef}), a MIDI-to-audio synthesis method that operates without MIDI annotations, thus circumventing the data collection challenges in supervised training. As shown in Fig.~\ref{fig:darefine-demo}, CoSaRef first concatenates note-level samples based on MIDI representation to generate synthetic audio. Then a deep audio generation model (\textit{DAGM}), based on diffusion model, is utilized to refine the synthetic audio into a realistic one. % pre-trained (or fine-tuned)
CoSaRef is inspired by zero-shot guided image editing methods~\cite{sdedit} and therefore does not require training on audio-audio or MIDI-audio paired datasets.
Moreover, CoSaRef allows for the control of the instrument timbre by adjusting MIDI and note-level sample input for the concatenative sampler. % thanks to the generalization capabilities of DAGMs. 
This type of timbre control is commonly found in the production workflow of digital audio workstations. Therefore, our approach can be easily integrated into this workflow without significantly altering musicians' typical design procedures.

Through experimental evaluation, CoSaRef first demonstrated its ability to generate realistic and expressive instrumental audio from MIDI input.
Simultaneously, the generated audio preserved the timbre of one-shot sample inputs while faithfully rendering the input MIDI score.
Furthermore, we evaluated CoSaRef using a DAGM fine-tuned on the target instrumental audio.
The results showed that CoSaRef could produce realistic audio with nuanced similarities to the target performance while remaining faithful to the input MIDI.
Despite not being supervised on MIDI annotation, CoSaRef's performance was superior to a state-of-the-art timbre-controllable method requiring MIDI-audio paired data for supervised training.
Audio samples and additional results are available on the demo page\footnote{\url{https://flymoons.github.io/midi-to-audio-demo/}}.

\section{Related works}
\label{sec:related-work}
Deep generative models have advanced MIDI-to-audio synthesis, persuing musical expressiveness~\cite{deepperformer, jonason2020control, castellon2020towards, midi-ddsp, spec-diffusion, kim2024guitar, maman2025taslp} and controllability~\cite{jonason2020control, midi-ddsp, castellon2020towards, Demerle2024ismir}.
MIDI-DDSP~\cite{midi-ddsp} enabled track-level detailed control over timbre and expression through model cascading and explicit expression modeling.
MIDI-DDSP represents the state of the art among the controllable MIDI-to-audio methods, surpassing other approaches~\cite{jonason2020control, castellon2020towards} in realism and timbre diversity of the output.
Some of the MIDI-to-audio works addressed the MIDI-audio paired data shortage by leveraging synthetic audio~\cite{spec-diffusion, kim2024guitar}. % or categorical performance conditioning~\cite{maman24perf}. 
This solution led to a decline in expressiveness and realism of the output audio.  
Moreover, these previous methods struggled with versatile track generations with various, user-intended timbres and performance styles. 
Timbre cloning from a reference track has been explored in the previous work~\cite{Demerle2024ismir}, which did not address the synthesis of realistic performances at the track level.
CoSaRef also employs a deep neural network but does not require MIDI-audio paired datasets for training, unlocking the ability to generate expressive tracks while preserving the timbre diversity.

CoSaRef draws inspiration from the zero-shot image editing methods using pre-trained text-to-image models~\cite{sdedit, ddpm-inversion}. Several studies in the audio domain investigated applying zero-shot editing approaches to audio style transfer~\cite{audioldm, mancusi2025latent} and more general audio/music manipulation~\cite{manor2024zeroshot, zhang2024musicmagus, zhang2024sdmuse}. 
It can be regarded that CoSaRef performs unsupervised musical timbre transfer~\cite{manor2024zeroshot, mancusi2025latent} tailored for MIDI-to-audio synthesis, transforming synthetic audio into a realistic one.

\section{Proposed method}
\label{sec:method}
Our proposed MIDI-to-audio synthesis method, CoSaRef, comprises two modules as shown in Fig.~\ref{fig:darefine-demo}: concatenative sampler and generative refinement module.
We dare not train the whole part of this architecture~---~the trainable part is DAGM in the generative refinement module. 
Therefore, CoSaRef does not require MIDI annotations for training.

\begin{figure}[t]
  \centering
  \hspace{0mm}
  \includegraphics[width=1.0\linewidth]{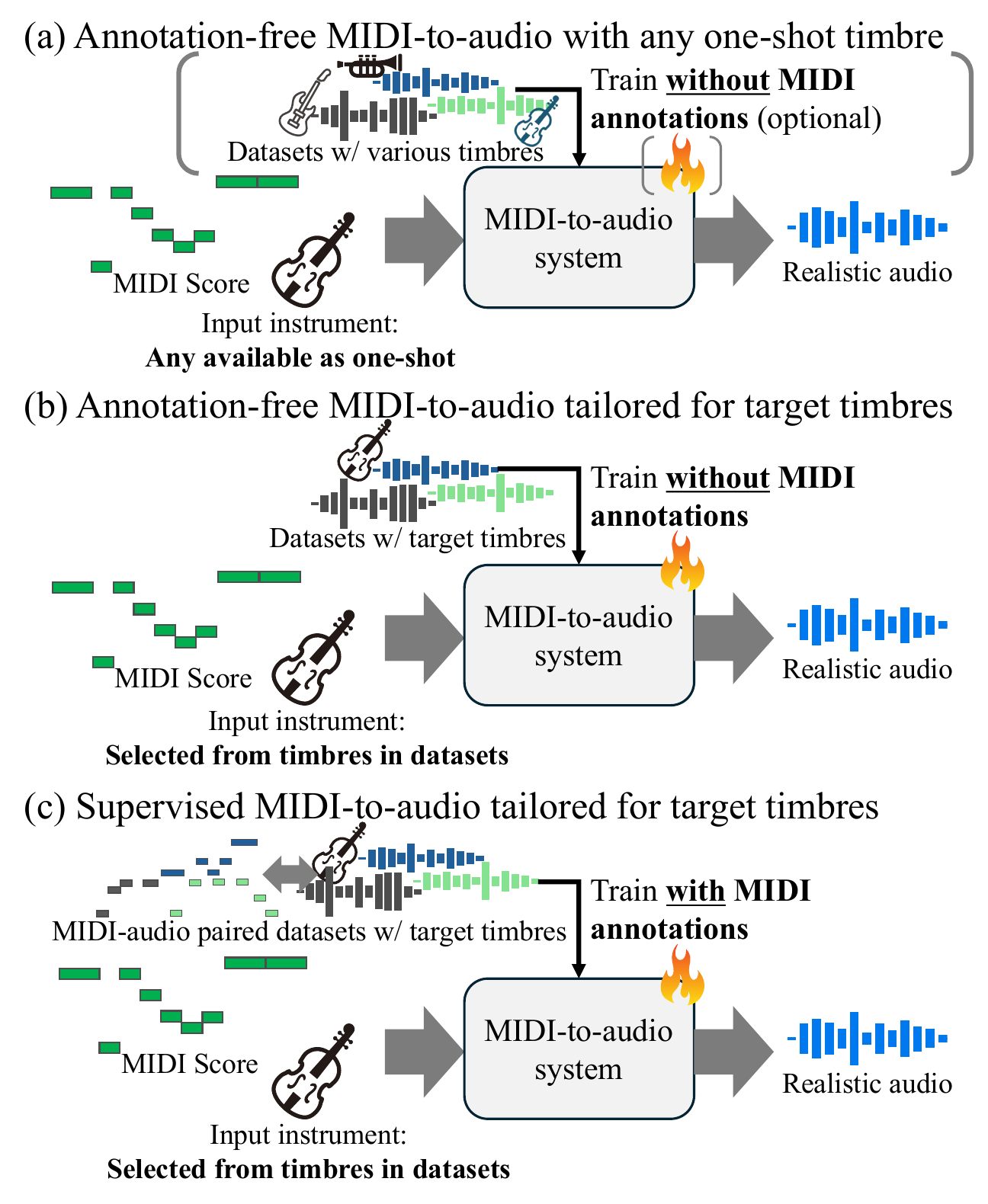}
\vspace{-8mm}
\caption{Diagrams of MIDI-to-audio problem settings. CoSaRef addresses the upper (a), (b) MIDI-to-audio settings, while conventional supervised learning-based methods address the lower (c) setting.}
\label{fig:midi2audio}
\vspace{-4mm}
\end{figure}

CoSaRef addresses two specific MIDI-to-audio synthesis challenges, depicted in Fig.~\ref{fig:midi2audio}. These settings remain unexplored in prior research using deep neural networks:

\noindent\textbf{Annotation-free MIDI-to-audio with any one-shot timbre.} In this problem setting, users are assumed to have a library of one-shot (i.e., note-level) samples for the desired timbres. 
This sample library is often accessible through a sample-concatenative synthesizer or similar means.
By using CoSaRef, realistic audio can be generated in any instrument timbre in that library, as shown in Fig.~\ref{fig:midi2audio}(a).

With this capability and the wide coverage of instrumental timbres in the DAGM training data, CoSaRef can generate realistic audio with various timbres that can be unseen in conventional supervised-based methods~\cite{midi-ddsp, spec-diffusion}.
This capability offers users enhanced control over timbre. % instrument timbre.

\noindent\textbf{Annotation-free MIDI-to-audio specialized in target timbres.} In this problem setting, users are supposed to have some reference performances with a target timbre.
CoSaRef, with the DAGM fine-tuned using these performance audio data, can generate realistic tracks that have the desired tonal characteristics.

The fine-tuning can be performed without MIDI annotations for audio, as shown in Fig.~\ref{fig:midi2audio}(b).  
Previous neural methods for expressive MIDI-to-audio synthesis employ a supervised training scheme, as depicted in Fig.~\ref{fig:midi2audio}(c), which poses a challenge due to the need for MIDI-annotated data collection.  
In contrast, CoSaRef can be developed using audio data without MIDI annotations, circumventing this data collection issue.

\subsection{Synthesizing tracks via Sample Concatenation}\label{ssec:concat-sampler}
The concatenative sampler takes a MIDI score as input, retrieves the corresponding note samples from pre-recorded libraries, and constructs a synthesized stereo track $\bm{s}_{\mathrm{syn}} \in \bR^{2\times T}$ of length $T$ samples.  
Given a musical note on the input MIDI score and a user-selected target instrument timbre, the simplest form of a concatenative sampler retrieves the corresponding note sample from the library based on its pitch and velocity.
The sample is then played back according to the specified duration and the sampler's envelope settings. 
A track is generated by applying this process to all notes in the MIDI input. 

The concatenative sampler provides high flexibility in controlling the output according to the music creator's intent, as traditionally practiced. 
There are no inherent restrictions on using extra MIDI input, such as sustain and expression (MIDI control change), or on selecting different concatenation methods of the sampler.
For instance, music creators can use either a simple concatenative sampler that merely plays back note samples or a more advanced concatenative sampler, selecting the optimal sample for each MIDI note at the track level.
Creators can also craft a detailed MIDI score as input to the sampler, emulating human performances.
These workflows align with established computer-based music production techniques familiar to trackmakers. 
Our approach ensures a high degree of control by adhering to these classic processes while leveraging the generative refinement module to produce a more realistic performance.

CoSaRef can generate a track with a diverse range of instrumental timbres using one-shot samples, as the concatenative sampler synthesizes a track based on note-level sample playback. 
When using a DAGM trained on tracks with a diverse range of instrumental timbres, CoSaRef can generate a realistic track without altering the timbre specified by the music creator in the sampler. 

\subsection{Diffusion-Based Refinement of Synthetic Audio}\label{ssec:method-refine}
Being provided a synthesized audio $\bm{s}_{\mathrm{syn}}$ from the concatenative sampler, the generative refinement module outputs a realistic audio $\bm{s}_{\mathrm{real}} \in \bR^{2\times T}$ faithful to $\bm{s}_{\mathrm{syn}}$.
It is constructed on a text-to-audio latent diffusion model (\textit{LDM}) as DAGM and the zero-shot audio editing methods~\cite{sdedit, manor2024zeroshot}. 

In the text-to-audio task, text-conditional LDMs aim at modeling the distribution of the compressed representation $\bm{z} \in \bR^{C\times (T/d)}$ of target audio conditioned by text representation $\bm{c}$, denoted as $p_{\mathrm{data}}(\bm{z}|\bm{c})$.
The representation is extracted by a pre-trained variational autoencoder (\textit{VAE})-based model with $C$ feature channels and a downsampling ratio of $d$.
Starting from the zero-mean isotropic Gaussian distribution $\mathcal{N}(\bm{0}, \sigma^2 \mathrm{I})$, LDMs gradually transform this distribution into the desired distribution by the denoising process. 
During the training, LDM first adds Gaussian noise to the input audio representation $\bm{z}_0\sim p_{\mathrm{data}}$ following the forward process below~\cite{song2021scorebased}:
\begin{align}
    \bm{z}_i = \alpha_i \bm{z}_0 + \sigma_i \bm{w},\ \bm{w} \overset{\mathrm{i.i.d.}}{\sim} \mathcal{N}(\bm{0}, \mathrm{I}),\  \label{eq:diff-forward}
\end{align}
where $i \in \left\{ 1, 2, \cdots, N\right\}$ denotes the index of each time step $t_i$ with $N$ denoted as the total step. $\alpha_i \in [0, 1)$ and $\sigma_i \in [0, \infty)$ represents the noise schedules to adjust the presence of input audio and noise in the update rule Eq.~(\ref{eq:diff-forward}), respectively. 
The reverse process is trained with a deep neural network $D(\bm{z}_i, t_i, \bm{c};\theta)$, named denoiser, to estimate the denoised VAE representation $\bm{z}_0$. Here we denote $\theta$ as the learnable parameters of the denoiser $D$.

The inference phase of LDM adopts a sampling strategy accompanied by a random noise input $\bm{w}_i\sim\mathcal{N}(\bm{0}, \mathrm{I})$, denoted as $S_{D}(\cdot, t_i;\bm{w}_i, \bm{c})$, to generate a representation $\hat{\bm{z}}_0$.  
During the inference, $S_{D}$ is operated recursively,
\begin{align}
    \hat{\bm{z}}_{i-1} = S_D (\hat{\bm{z}}_i, t_i; \bm{w}_i, \bm{c}), \hat{\bm{z}}_N \equiv \bm{z}_N, \label{eq:diff-backward}
\end{align}
to gradually invert the forward process Eq.~(\ref{eq:diff-forward}) and eventually get a sampled output representation $\hat{\bm{z}}_0$.

Based on this LDM, we repurpose existing image and audio editing methods~---~specifically, SDEdit~\cite{sdedit} and ZETA~\cite{manor2024zeroshot}~---~as back-end techniques for the refinement:

\noindent\textbf{SDEdit. }
\begin{figure}[t]
  \centering
  \hspace{0mm}
  \includegraphics[width=1.0\linewidth]{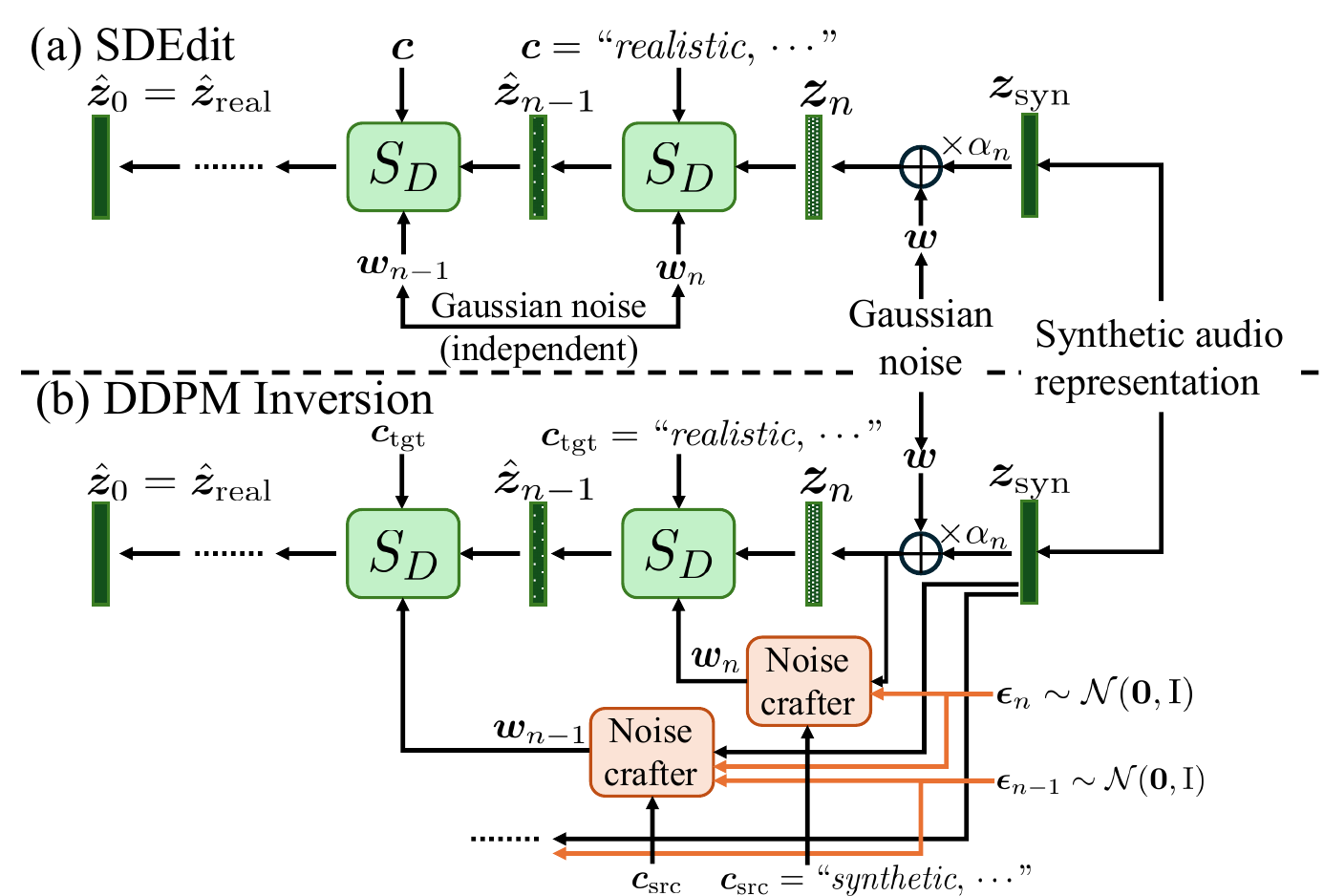}
\vspace{-7mm}
\caption{Diagrams of the proposed generative refinement modules in CoSaRef using SDEdit~(a)~\cite{sdedit} and DDPM inversion~(b)~\cite{ddpm-inversion}. 
$S_D$ denotes the sampling strategy.
}
\label{fig:darefine-refinemodule}
\vspace{-3mm}
\end{figure}
With SDEdit, this module first runs the forward process in Eq.~(\ref{eq:diff-forward}), using $p_{\mathrm{data}}$ as the distribution of the synthetic audio domain, with the time index $i = n < N$. The module reinterprets the distribution $p(\bm{z}_n|\bm{c})$ as a Gaussian-blurred distribution of the realistic audio domain. The module then performs reverse sampling in Eq.~(\ref{eq:diff-backward}) from the intermediate time step $t_n$ to get the realistic audio representation $\hat{\bm{z}}_{\mathrm{real}}$, as depicted in Fig.~\ref{fig:darefine-refinemodule}(a). 

\noindent\textbf{ZETA.} 
Unlike SDEdit, ZETA employs DDPM inversion~\cite{ddpm-inversion} to run reverse sampling with crafted noise in which the information of the synthetic audio representation $\bm{z}_{\mathrm{syn}}$ is strongly imprinted.
In the generative refinement module depicted in Fig.~\ref{fig:darefine-refinemodule}(b), DDPM inversion designs an edit-friendly noise $\bm{w}_i$ directly from $\bm{z}_{\mathrm{syn}}$, normal noise $\bm{\epsilon}_i$, and a text $\bm{c}_{\mathrm{src}}$ describing synthetic audio $\bm{s}_{\mathrm{syn}}$ for each step index $i$.
Then DDPM inversion runs the reverse process from the index $n$ via Eq.~(\ref{eq:diff-backward}) with the crafted $\bm{w}_{i}$ and the target text prompt $\bm{c}_{\mathrm{tgt}}$ describing realistic audio, to obtain the representation of target realistic audio $\hat{\bm{z}}_{\mathrm{real}} \equiv \hat{\bm{z}}_0$.

As a previous study~\cite{manor2024zeroshot} reported, we expect that compared to SDEdit, ZETA can refine the synthetic audio while preserving global features such as melody and timbre. 

\vspace{-0.5mm}
\subsection{Fine-tuning of DAGMs in CoSaRef}\label{ssec:method-ft}
Fine-tuning DAGM on proper audio datasets without MIDI annotations enhances the realism of CoSaRef’s outputs and improves their alignment with the desired styles. 
First, in MIDI-to-audio synthesis with one-shot timbres, as shown in Fig.~\ref{fig:midi2audio}(a), fine-tuning DAGM on general realistic audio is expected to improve the realism of CoSaRef’s outputs. 
Pre-trained DAGMs were often trained on a mix of realistic or synthetic instrumental phrases and one-shot programmed samples.
Fine-tuning can align them with the domain of realistic audio that plays phrases.
In addition to the text-conditioning guidance described in Sec.~\ref{ssec:method-refine}, this adaptation helps refine the generated audio to be more natural.
Second, fine-tuning DAGMs on specific instrumental performances enables CoSaRef to better capture the nuances of the target performance style in the generated audio. 
In MIDI-to-audio synthesis with target timbres, depicted in Fig.~\ref{fig:midi2audio}(b), this fine-tuning allows CoSaRef to generate more realistic tracks with timbres and expression styles more closely matching the target performances than those produced using pre-trained DAGMs.

During the fine-tuning, the text conditioning $\bm{c}$ is needed for each input audio. This conditioning can facilitate the text guidance of diffusion sampling in the generative refinement module. 
For each audio entry in the fine-tuning dataset, we retrieve the instrument type from the metadata and automatically insert it into a predefined template to obtain text conditioning.

\section{Experimental Evaluation}
\label{sec:exp}
In this experiment, we assessed whether CoSaRef effectively generated realistic and expressive instrumental audio that accurately followed the input MIDI score and desired timbres.
First, we compared the performance of CoSaRef to the concatenative sampler without generative refinement when synthesizing audio from MIDI and one-shot timbre inputs.
Second, we evaluated whether CoSaRef, with the fine-tuned DAGM, could generate realistic instrumental performances that matched a specific target performance style.
For this evaluation, we compared the performance of this type of CoSaRef with MIDI-DDSP~\cite{midi-ddsp}. 
MIDI-DDSP is the state-of-the-art neural MIDI-to-audio synthesis method that allows flexible control of the output audio by editing MIDI input and timbres.
Note that MIDI-DDSP was trained on a MIDI-audio paired dataset, addressing supervised MIDI-to-audio synthesis in Fig.~\ref{fig:midi2audio}(c).

In this experiment, we evaluated the automatic generation of instrumental audio given a MIDI input. 
As discussed in Sec.~\ref{ssec:concat-sampler}, CoSaRef can reflect the musical intentions embedded in carefully designed MIDI input and timbres.
However, this experiment focused on the automatic generation of audio, not the manual design of the input MIDI and timbre.

We evaluated CoSaRef on the generation of single, monophonic instrumental tracks.
Subjective and objective evaluations were conducted to measure the performance of MIDI-to-audio synthesis methods in this experiment.

\vspace{-1mm}
\subsection{Setup}
\label{ssec:exp-setup}
\textbf{Target realistic audio. } We used individual instrument performances from the URMP dataset~\cite{urmp} as the target for realistic instrumental audio. The sampling frequency of the audio was converted to $44.1\, \mathrm{kHz}$. %Instead of the precise note on/offset alignment from the ground-truth audio, 
We used the MIDI score sheet as MIDI input for the MIDI-to-audio synthesis methods. For the train/test split, we followed the experimental settings from the MIDI-DDSP paper~\cite{midi-ddsp}.

\noindent\textbf{Concatenative sampler.} We used the train split of NSynth dataset~\cite{nsynth2017} as the sample library of concatentive sampler in Fig.~\ref{fig:darefine-demo}. 
For the synthetic audio generation, we manually selected a sample from the NSynth dataset for each URMP instrument type based on the internal listening test. 
To ensure that concatenative sampler functions as a typical sampler that plays sustaining, natural-sounding notes, we fixed the parameters of its ADSR envelope to an attack of $5\, \mathrm{ms}$, a sustain of $100\, \%$, and a release of $200\, \mathrm{ms}$.
% As a typical concatenative sampler playing sustaining and natural-sounding notes, attack, sustain, and release of the sampler was fixed to $5 \, \mathrm{ms}, 100 \, \%$, and $100\, \mathrm{ms}$. 
Both the attack and release had the linear property. 
% In rendering notes, the sample values after $90\, \%$ of the note duration were set to decay exponentially. 
The sample concatenation was processed in the sampling frequency of $16\, \mathrm{kHz}$, and the output audio was resampled to $44.1 \, \mathrm{kHz}$.

\noindent\textbf{Generative refinement module.} We adopted Stable Audio Open~\cite{stableaudioopen} (\textit{SAOpen}) as DAGM in the generative refinement module. 
The timing conditioning~\cite{stableaudio} was fixed to $seconds\_start=0 \, \mathrm{s}$ and $seconds\_total=47.0 \, \mathrm{s}$.
For the text conditioning, CoSaRef with SDEdit used $\bm{c} = \,$``\textit{Solo, realistic, }\textrm{instrument\_type},\textit{ classical, well-recorded, professional}'', and CoSaRef with ZETA used $\bm{c}_{\mathrm{src}} = \,$``\textit{synthetic, }\textrm{intrument\_type}'' and $\bm{c}_{\mathrm{tgt}} = \,$``\textit{realistic, }\textrm{intrument\_type}'', with instrument\_type replaced by the corresponding instrument category in URMP.

% For the text conditioning style was fixed to ``\textit{Solo, realistic, }\textrm{instrument\_type},\textit{ classical, well-recorded, professional}'', with instrument\_type replaced by the corresponding instrument category in URMP.
For the sampling strategy $S_{D}$, we applied second-order multistep DPM-solver++~\cite{dpmsolver-pp}.
Classifier-free guidance~\cite{ho2021classifierfree} was applied to inject conditioning inputs.
Table~\ref{tab:hyperparams-cosaref} shows the hyperparameters of the generative refinement module, following the notation of k-diffusion~\cite{k-diffusion} settings. %of $\sigma_{\min} = 0.05$, $\rho = 1$, and the total step $N = 250$, following the paper~\cite{stableaudioopen}. 
\begin{table}[t]
    \centering
    \caption{Hyperparameters of sampling strategy $S_D$ for CoSaRef's generative refinement module.} % following the notation of k-diffusion~\cite{k-diffusion} settings.}
    \label{tab:hyperparams-cosaref}
    % \resizebox{\linewidth}{!}{
    \begin{tabular}{r|cc}
        \toprule
        Back-end method for refinement & SDEdit & ZETA \\
        \midrule
        Initial noise magnitude $\sigma_{\mathrm{max}}$ & $16$ & $500$ \\
        $\sigma_{\mathrm{min}}$ & $0.05$ & $0.3$ \\
        scale of classifier-free guidance & $7.0$ & $4.0$ \\
        \# of diffusion steps $N$ & $250$ & $200$ \\
        Intermediate step index $n$ & $150$ & $70$ \\
        \bottomrule
    \end{tabular}
    % }
    \vspace{-4mm}
\end{table}
These hyperparameters were set based on the SAOpen implementation~\cite{stableaudioopen} and preliminary experiments.

Since SAOpen generates audio up to 47 seconds, the refinement of module cannot directly handle $\bm{s}_{\mathrm{syn}}$, which is typically longer. 
To address this, $\bm{s}_{\mathrm{syn}}$ was split into 47-second chunks, processed individually, and then concatenated with a 1000-sample overlap to generate the output.

\noindent\textbf{Fine-tuning of DAGM.} We tested three DAGMs: 1) pre-trained SAOpen using the official weights\footnote{\url{https://huggingface.co/stabilityai/stable-audio-open-1.0}}, 2) SAOpen fine-tuned on multiple datasets (\textit{Multi-data}) excluding URMP, and 3) SAOpen fine-tuned on the URMP train split. 
Model 2) was developed for generating realistic audio with various timbres, addressing MIDI-to-audio synthesis with one-shot samples.
Model 3) was designed to handle MIDI-to-audio synthesis with target timbres, refining the output to capture the nuances of URMP styles and timbres.

We fine-tuned the Diffusion Transformer (\textit{DiT}) module in SAOpen and froze the remaining parts.
For fine-tuning model 2), we collected Solos~\cite{montesinos2020solos}, Medley-solos-DB~\cite{medley-solosdb}, PHENICX-Anechoic~\cite{phenicx}, and MAESTROv3~\cite{maestro}. The initial model weights were set to those of the official SAOpen in all fine-tuning experiments. 
%The initial checkpoint was set to the official weights of SAOpen in all fine-tuning experiments. 
During the fine-tuning, % the Diffusion Transformer module was trained on an NVIDIA A100 GPU with a batch size of 8. 
The DiT module in model 2) was trained for 62000 steps with a batch size of 64, while that in model 3) was trained for 8000 steps with a batch size of 8.
The other training configurations followed the SAOpen implementation. 
% The timing conditioning was fixed to $seconds\_start=0 \, \mathrm{s}$ and $seconds\_total=47.0 \, \mathrm{s}$. 
The text conditioning attached to each audio had the same style as $\bm{c}$ in CoSaRef inferences using SDEdit.

\subsection{Evaluation Schemes}\label{ssec:eval-scheme}
\textbf{Compared methods.} Four MIDI-to-audio synthesis methods were developed and tested. \textit{Concat} denotes the concatenative sampler without any refinement. \textit{CoSaRef (SDEdit)} and \textit{CoSaRef (ZETA)} correspond to the proposed CoSaRef itself using SDEdit and ZETA as a refinement back-end, respectively. \textit{MIDI-DDSP}~\cite{midi-ddsp} method generates the expressive audio directly from the MIDI score sheet input. MIDI-DDSP model weights were obtained by training on URMP\footnote{available on \url{https://github.com/magenta/midi-ddsp}.} that was treated as a MIDI-audio paired dataset here. In this experiment, the output audio of MIDI-DDSP was resampled from $16 \, \mathrm{kHz}$ to $44.1 \, \mathrm{kHz}$.

\noindent\textbf{Metrics of objective evaluation.} We used Fr\'echet audio distance~\cite{FAD} (\textit{FAD}) and F1 Score of MT3\footnote{weights available on \url{https://github.com/magenta/mt3}}~\cite{mt3} transcription (\textit{F1}) for the performance evaluation. FAD aims at measuring the audio perceptual similarity to the reference audio set, and F1 captures the output audio's faithfulness to the input MIDI score. 

We adopted two FAD measures for assessing different performances in MIDI-to-audio synthesis.
First, we computed \( \text{FAD} \) using CLAP-LAION~\cite{laionclap2023} embeddings, taking the ground truth URMP recordings as the reference set, named as $\mathrm{FAD}_r$. 
This metric assessed the realism of the audio outputs and their closeness to the nuances of the target performance.
Second, we computed \( \text{FAD} \) using Encodec’s continuous embeddings~\cite{defossez2022highfi} with a 48 kHz input, taking the output of the concatenative sampler as the reference set, named as $\mathrm{FAD}_r$. 
This metric specifically measured how well CoSaRef preserved the timbre distribution~\cite{fadtk, mancusi2025latent} of the synthetic audio after refinement.
These FAD were calculated using \texttt{fadtk}~\cite{fadtk} implementation.

\noindent\textbf{Subjective evaluation.} We conducted a mean opinion score (\textit{MOS}) test to evaluate MIDI-to-audio synthesis with one-shot samples in terms of realism and human-like expressiveness.
In this test, participants rated the instrumental audio generated by MIDI-to-audio methods on a 5-point MOS scale, where 1 was ``very synthetic'' and 5 was ``very realistic.'' % evaluating its realism and human-likeliness.
We collected ratings for ground truth recordings as well as for the audio outputs of Concat, CoSaRef (SDEdit), CoSaRef (ZETA), and MIDI-DDSP.
In this test, only CoSaRef using pre-trained DAGM was evaluated.
% we conducted a mean opinion score (MOS) test to assess various methods, including CoSaRef using DAGM 1).

We conducted an A/B preference test to evaluate CoSaRef’s performance in MIDI-to-audio synthesis with target timbres. 
This test assessed the realism of audio generated by CoSaRef (SDEdit) using DAGM 3), comparing it with the audio produced by MIDI-DDSP.
Participants listened to the audio generated by the two methods and chose which one sounded more realistic and human-like.
% To mitigate the influence of differences in timbre and performance phrases, we presented audio samples generated by the two methods in a manner that ensured a fair comparison. 
% Specifically, we stated the corresponding instrument for each audio sample and synchronized the 10-second excerpts between the two methods. 
Note that A/B preference test for CoSaRef (ZETA), which showed lower performance according to the objective evaluation results presented in Sec.~\ref{ssec:results}, was not conducted.

\subsection{Results and Discussion}\label{ssec:results}
%The evaluation results on URMP are shown in Table~\ref{tab:results}.
\subsubsection{MIDI-to-audio with any one-shot timbre}\label{sssec:result-oneshot}
We first confirmed that CoSaRef was effective for generating realistic audio with MIDI input and one-shot samples through the objective evaluation result shown in Table~\ref{tab:result-main-obj}.
\begin{table}[t]
    \centering
    \caption{
    Results of the objective evaluation of MIDI-to-audio synthesis methods. ``DAGM'' column describes how the DAGM in CoSaRef was fine-tuned, specifically indicating the data used for training. \textbf{Timbre} column presents the timbre similarity metric relative to the output of the concatenative sampler. \textbf{Score} column provides the adherence metric to the MIDI input score sheet.
    The worst value of $\mathrm{FAD}_t$ (${}^\ast$) serves as a reference, as MIDI-DDSP does not aim to generate audio with one-shot timbres from NSynth.
    % Result of objective evaluation on MIDI-to-audio synthesis methods. ``DAGM'' column describes how the DAGM in CoSaRef was fine-tuned, especially indicating the data for training. \textbf{Timbre} column shows the timbre similarity metric to the Concat's output. \textbf{Score} column shows the adherence metric to the MIDI input \textbf{score}.
    }
    \label{tab:result-main-obj}
    \resizebox{\linewidth}{!}{
    \begin{tabular}{c|c|ccc}
        \toprule
       \multirow{2}{*}{Method} & \multirow{2}{*}{DAGM} & \textbf{Realism} & \textbf{Timbre} & \textbf{Score} \\
       & & $\mathrm{FAD}_r\downarrow$ & $\mathrm{FAD}_t\downarrow$ & F1$\uparrow$ \\\midrule
       Concat & -  & 0.712 & - & 0.325 \\\midrule
       \multicolumn{5}{l}{\textbf{Annotation-free MIDI-to-audio with one-shot timbres}} \\\midrule
       \multirow{2}{*}{\begin{tabular}{c} CoSaRef\\(SDEdit)\end{tabular}} & Pre-trained & 0.548 & \textbf{1.560} & 0.354 \\
       & Multi-data & 0.388 & 4.633 & 0.256 \\\midrule
       \multirow{2}{*}{\begin{tabular}{c} CoSaRef\\(ZETA)\end{tabular}} & Pre-trained & 0.380 & 1.742 & \textbf{0.375} \\
       & Multi-data & \textbf{0.355} & 2.301 & 0.281 \\\midrule
       \multicolumn{5}{l}{\textbf{Annotation-free MIDI-to-audio with target timbres}} \\\midrule
       \begin{tabular}{c} CoSaRef\\(SDEdit)\end{tabular} & URMP & \textbf{0.235} & 5.183 & \textbf{0.412} \\\midrule
       \begin{tabular}{c} CoSaRef\\(ZETA)\end{tabular} & URMP & 0.348 & \textbf{0.987} & 0.372 \\\midrule
       \multicolumn{5}{l}{\textbf{Supervised MIDI-to-audio with target timbres}} \\\midrule
       \begin{tabular}{c}
            MIDI- \\ DDSP
       \end{tabular} & - & 0.526 & (20.291)${}^\ast$ & 0.323 \\\bottomrule
    \end{tabular}
    }
    \vspace{-3.5mm}
\end{table}
The result showed an improvement in $\mathrm{FAD}_r$ across all CoSaRef-based methods compared to Concat.
Additionally, using ZETA for refinement improved $\mathrm{FAD}_r$ consistently, compared to using SDEdit.
Similarly, using ``Multi-data'' DAGM for refinement improved $\mathrm{FAD}_r$ compared to pre-trained DAGM.
% Similarly, using ``Multi-data'' DAGM for refinement improved $\mathrm{FAD}_r$ compared to using pre-trained DAGM.
Furthermore, CoSaRef with the pre-trained DAGM achieved a higher F1 than that of Concat.
CoSaRef with ``Multi-data'' DAGM showed a lower F1 than that of Concat, while the refinement by ZETA improved it.
These results indicate that CoSaRef, especially with the refinement by ZETA, enhanced realism and human-likeness, and faithfully played the input MIDI score in MIDI-to-audio synthesis using one-shot samples, outperforming the simple sample concatenation approach.
Fine-tuning the DAGM on appropriate audio data improved the realism of the generated audio, but it compromised the faithful MIDI rendering, suggesting that the DAGM overfitted to phrases in the datasets.

The objective evaluation also showcased whether the generative refinement module in CoSaRef could preserve the timbre of input synthetic audio in the generated outputs.
According to Table~\ref{tab:result-main-obj}, $\mathrm{FAD}_t$ of CoSaRef performing MIDI-to-audio with one-shot timbres was consistently smaller than that of MIDI-DDSP or CoSaRef with SDEdit using the ``URMP'' DAGM, which specialized in generating URMP-style audio.
Furthermore, when CoSaRef was run with ``Multi-data'' DAGM, applying ZETA in the refinement improved $\mathrm{FAD}_t$. 
This improvement was larger than the $\mathrm{FAD}_t$ degradation observed when the pretrained DAGM was used.
These results indicate that CoSaRef, particularly with ZETA, effectively preserves timbre fidelity while enhancing the realism of generated audio. 
We encourage readers to listen to generated audio samples in the supplementary materials, showcasing the CoSaRef performance in timbre fidelity.

Additionally, Table~\ref{tab:result-main:mos} presents the results of the MOS test evaluating the realism of audio generated by MIDI-to-audio synthesis methods.
\begin{table}[t]
    \centering
    \caption{MOS test result about the realism of generated audio, with $95\%$ confidence intervals. 
    % All the CoSaRef-based methods listed in this table utilized pre-trained DAGM for refinement.
    The \textbf{bold} score highlights the best performance in MIDI-to-audio methods with one-shot samples and its significant improvement ($p = 0.035$) compared to Concat.
    MIDI-DDSP (${}^\ast$) achieved a higher MOS, serving as a reference, since it did not target MIDI-to-audio with any one-shot timbre.}
    \label{tab:result-main:mos}
    % \resizebox{0.7\linewidth}{!}{
    \begin{tabular}{c|c}
        \toprule
         Method & Realism MOS \\\midrule
         Concat & $2.534\pm0.203$\\
         CoSaRef (SDEdit) & $2.570\pm0.204$\\
         CoSaRef (ZETA) & $\mathbf{2.837\pm0.192}$\\\midrule
         MIDI-DDSP & $(3.267\pm0.203)^\ast$\\\midrule
         Ground truth & $3.993\pm0.197$ \\\bottomrule
    \end{tabular}
    % }
    \vspace{-3mm}
\end{table}
The MOS test involved 29 participants, each of whom evaluated 29 audio segments with a length of 10 seconds.
% The first four questions presented two segments of Concat and the ground truth. 
Responses from two participants were excluded, rating Concat higher than the ground truth in dummy questions. % , were excluded, and the responses of 27 participants were investigated. 
% As shown in Table~\ref{tab:result-main:mos}, 
The results reveal that CoSaRef, with pre-trained DAGM and refinement by ZETA, generated significantly more realistic audio than Concat. % based on a two-sided t-test.

% Note that while MIDI-DDSP achieved a significantly higher realism MOS than CoSaRef (ZETA) in the test, CoSaRef (ZETA) had a lower $\mathrm{FAD}_r$ value.
% This result may be attributed to CoSaRef preserving timbres derived from NSynth, which could have led participants to perceive its output as monotonous and synthetic in timbre. % across phrases.

\subsubsection{MIDI-to-audio specialized in target timbres}
Table~\ref{tab:result-main-obj} also presents the objective evaluation results for MIDI-to-audio synthesis methods targeting the URMP performance style. 
% The evaluation investigated the realism of the generated audio and its similarity to the URMP style.
The table shows that CoSaRef (SDEdit) using ``URMP'' DAGM improved $\mathrm{FAD}_r$ and F1 compared to MIDI-DDSP. 
This result demonstrates that even CoSaRef, developed without MIDI annotations, outperformed MIDI-DDSP, which relies on a MIDI-audio paired dataset, in terms of realism, faithfulness to MIDI input, and style adherence to references.

On the other hand, Table~\ref{tab:result-main-obj} reveals that CoSaRef (ZETA) with the ``URMP'' DAGM led to a deterioration in $\mathrm{FAD}_r$ compared to CoSaRef (SDEdit), while resulting in a lower $\mathrm{FAD}_t$ value.
Through this observation and an internal listening test, we found that CoSaRef (ZETA) failed to refine the synthetic audio and generated audio that sounded almost the same as the synthetic one.
% Therefore, ``URMP'' DAGM in this study was suitable for refinement with SDEdit but not with ZETA.
This result and the DAGM overfitting observed in Sec.~\ref{sssec:result-oneshot} suggest that fine-tuning DAGMs requires careful hyperparameter tuning, depending on the refinement back-end method.
Developing a method that operates robustly across various back-end methods for refinement remains future work.
% Developing a refinement method that operates robustly across various back-end methods remains future work.
% This observation and DAGM overfitting observed in Sec.~\ref{sssec:result-oneshot} suggest that DAGM fine-tuning requires careful hyperparameter tuning depending on the refinement back-end method.

Additionally, we collected 434 responses from 22 evaluators for the A/B preference test described in Sec.~\ref{ssec:eval-scheme}. 
\begin{figure}[t]
  \centering
  \hspace{0mm}
  \includegraphics[width=1.0\linewidth]{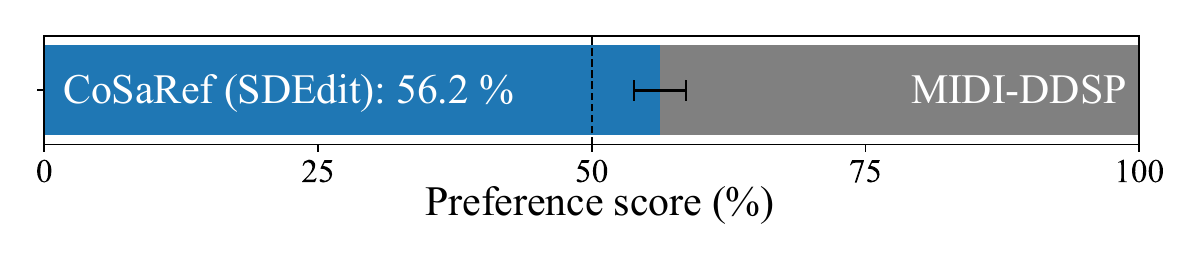}
\vspace{-9mm}
\caption{Preference score from A/B test for CoSaRef (SDEdit) using ``URMP'' DAGM and MIDI-DDSP, with $95\%$ confidence intervals indicated by the error bar.}
\label{fig:ab-test-result}
\vspace{-4mm}
\end{figure}
The result shown in Fig.~\ref{fig:ab-test-result} demonstrated a significant preference for its realism over MIDI-DDSP ($p = 0.009$).
% Additionally, the hyperparameters used during inference might not be appropriate for this case.

\section{Conclusion}
\label{sec:conclusion}

We proposed CoSaRef, a MIDI-to-audio synthesis method that does not require MIDI annotations on audio for development.
% We proposed CoSaRef, a MIDI-to-audio synthesis method requiring audio-only datasets for development. 
CoSaRef first uses a concatenative sampler with a sample library to generate synthetic audio from a MIDI input score.
% CoSaRef first uses a concatenative sampler with a sample library to obtain synthetic audio from a MIDI input score. 
A diffusion-based generative model then refines the synthetic audio, transforming it into more realistic sound by running the reverse process from an intermediate time step. 
This approach addresses the data shortage challenge faced by previous MIDI-supervised methods, enabling enhanced expressiveness and timbre diversity with audio datasets without MIDI annotations.
Additionally, CoSaRef allows music creators to control the output timbre by selecting one-shot samples.
% Then, a diffusion-based generative model transforms the synthetic audio into more realistic audio by running the reverse process from an intermediate time step. This approach circumvents the challenge of data shortage existing in previous methods based on MIDI-supervised training, which makes it possible to enhance the expressiveness with datasets only containing audio. Additionally, CoSaRef enables music creators to control the output timbre by selecting one-shot samples. 
Experimental evaluation demonstrated that even without MIDI annotations, CoSaRef outperformed conventional methods in generating realistic tracks that faithfully play MIDI input, while preserving timbral details or capturing the nuances of target performance styles.
Future work should investigate CoSaRef on generating polyphonic, multi-track audio.
Furthermore, future research should pursue a lightweight, real-time implementation of CoSaRef to facilitate its practical use in music composition workflows.

% For BibTeX users:
\bibliographystyle{IEEEbib}
\bibliography{refs}

\end{document}